# Automatic Classification of Laser Peening Quality Using Acoustic Signals

Bohumil Kolář[1,*], Jan Kočí[1], Michal Kotek[1], Tomáš Martinec[1], Ivan Mašín[1]

[1]*Technical University of Liberec, Institute for Nanomaterials, Advanced Technology and Innovation, Department of Process Modelling and Artificial Intelligence, Liberec, Czech Republic*

[*]Corresponding author: bohumil.kolar@tul.cz

**Abstract**: Laser Shock Peening increases the fatigue life of metallic components by introducing beneficial compressive residual stresses. To achieve the desired effect, each individual laser pulse must be delivered correctly. Laser Shock Peening quality is typically verified by destructive and time-consuming residual stress measurements or by subjective operator judgement, which is non-objective and unsuitable for continuous in-line control. We propose a simple, low-cost and robust method based on the analysis of the acoustic response that automatically classifies individual laser pulses as defect-free or defective. We show that the acoustic response captured by a low-cost microphone carries sufficiently informative signatures to reliably distinguish correct from incorrect impacts and enables quality control at the level of single pulses. The method provides a non-destructive and objective route to real-time monitoring of Laser Shock Peening, with the potential to increase process reliability and support industrial deployment of this technology without the need for subsequent destructive measurements.

## 1. Introduction

Laser Shock Peening (LSP) improves the fatigue resistance of metallic parts by introducing compressive residual stresses [1–7]. For stable results, it is necessary to control the correct execution of every single pulse. However, common quality control procedures are destructive, slow or subjective, which prevents reliable in-line process control [2,3].

Research is focused on the relationship between LSP parameters and the resulting mechanical properties [1–5,7,8]. On-line monitoring methods (acoustic emission, vision, machine learning) are emerging, but their industrial deployment remains complex and limited [9–19].

Current approaches often require expensive sensors or sensitive camera systems, which complicates integration into production [9–14,19]. In addition, acoustic signals are strongly dependent on the measurement conditions, so a simple and robust solution is needed.

We propose (see Fig. 1) a low-cost method for real-time classification of individual LSP pulses using the acoustic response and features extracted from the audible band. The O*K*/*NOT OK* classification is performed using a Random Forest model offering straightforward options for industrial integration [8,20].

- Real-time diagnostics of the quality of individual pulses.
- Simple and inexpensive implementation using a standard microphone.
- Fully defined and tested pipeline suitable for industrial deployment.

We present the data acquisition, preprocessing, and feature extraction steps, followed by classification and an analysis of its robustness for practical industrial use.

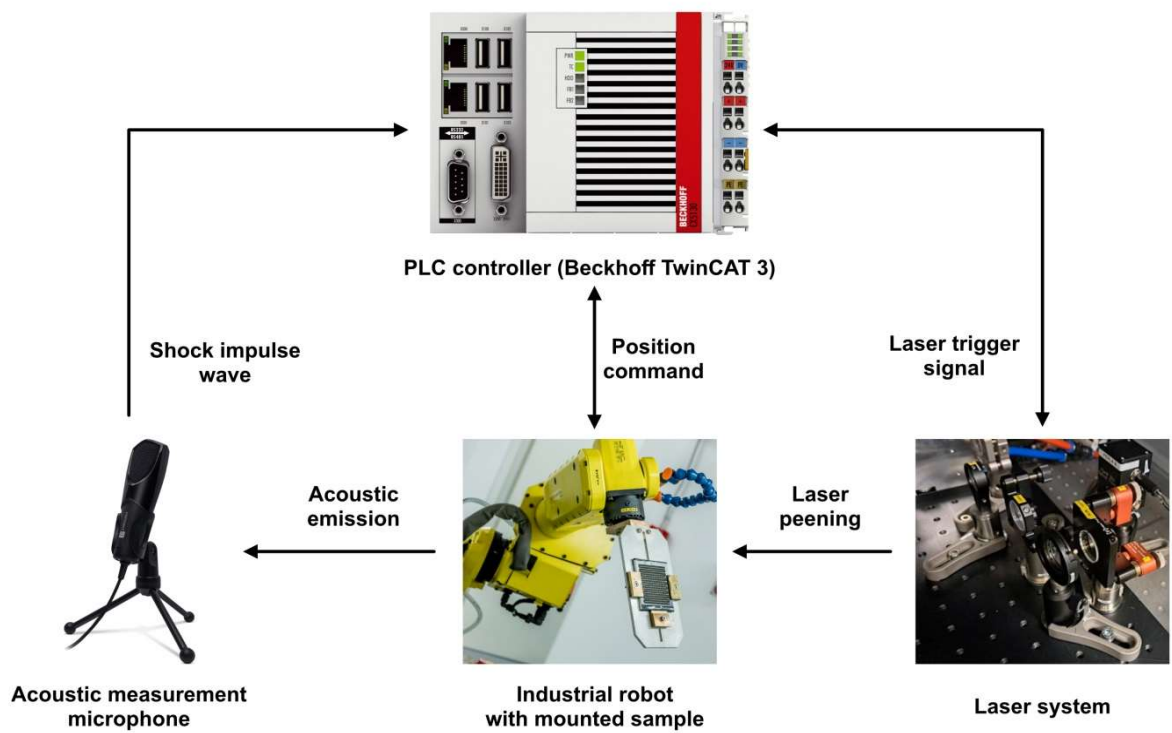

Figure 1: Schematic diagram of the proposed laser peening quality classification workflow.

## 2. Acoustic capture of LSP pulses

The acoustic signal was recorded with a condenser USB microphone with a sampling frequency of 44 kHz in mono mode. The acoustic data used in this work were acquired at the HiLASE laser centre from the LSP process on aluminium alloy AA2024 in the form of plates with dimensions 100 × 100 × 5 mm.

Shock pulses were generated by an Nd:YAG laser (Litron LPY ST 7875-10 2HG) with a wavelength of 1064 nm, a repetition rate of 1 Hz and a pulse energy of 1.5 J. To protect the surface from oxidation and ensure more stable energy transfer, a layer of vinyl tape was applied to the sample surface.

The LSP process was studied under two experimental conditions:

- **PROCESS OK** - with a water layer, **99** samples,
- **PROCESS NOT OK** - without a water layer, **99** samples.

The long LSP acoustic recording was segmented using amplitude-based detection. The normalized signal was converted to absolute amplitude and smoothed using a moving average with a 5 ms window. The detection threshold was defined as the mean plus three times the standard deviation of the envelope. Above-threshold regions were merged into individual pulses with a minimum spacing of 300 ms. For each pulse, a 0.4 s segment (0.1 s before and 0.3 s after the detected maximum) was then extracted.

## 3. Signal Processing and Analysis

Individual LSP pulses were subjected to time-frequency analysis using a combination of numerical methods. Basic characteristics were obtained by computing the root mean square (RMS) value and the spectral centroid using the fast Fourier transform (FFT). To detect pulses and describe their dynamics, an envelope analysis with a Gaussian window was employed. The entire process was visualized using time-domain waveforms, spectra and spectrograms, which enabled identification of characteristic properties of individual pulses.

**Time-domain waveform.** The time-domain waveform (see Fig. 2) complemented by its envelope represents the moment of the main impact of the laser pulse. The envelope also enables identification of the onset of the first pressure wave before the signal, as well as the subsequent oscillations caused by resonance, reflections, and cavitation.

**Detailed zoom of the main pulse.** A detailed zoom of the main pulse (see Fig. 3) makes it possible to observe the steepness of the shock wave and the subsequent cavitation phenomena, which appear as local maxima (peaks). Given the sampling frequency of 44 kHz, these phenomena occur within a very short time and partially overlap in the frequency domain. For pulses of the OK class, an average of 1.11 cavitation peaks was detected, whereas for pulses of the NOT OK class only 0.71. Peaks were defined as local maxima of the envelope exceeding 55 % of its amplitude in a specified time interval after the main pulse.

**Amplitude spectrum.** The amplitude spectrum (see Fig. 4) confirms the broadband character of the shock wave, with dominant spectral energy around 5 kHz. The frequency profile, together with the spectral centroid, provides quantitative information about the "sharpness" of the pulse. A higher centroid value corresponds to a greater presence of high-frequency components and therefore a more intense pressure wave [3,9,13].

**Time-frequency analysis.** The spectrogram (see Fig. 5) provides a time-frequency view that clearly shows the transition from the broadband main pulse to the gradual decay of high frequencies. At later times, low-frequency components dominate, corresponding to mechanical ringing of the structure and residual cavitation. The interpretation of the individual time intervals is summarized in the attached table (see Table 1).

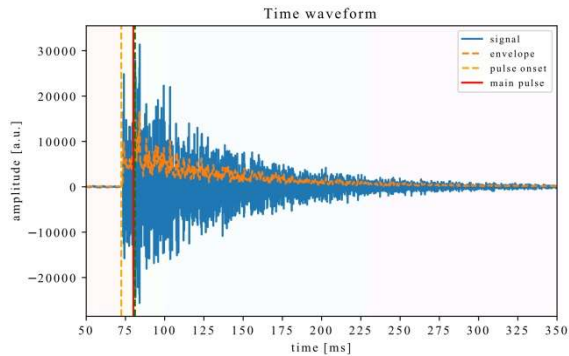

Figure 2: Time-domain waveform (OK)

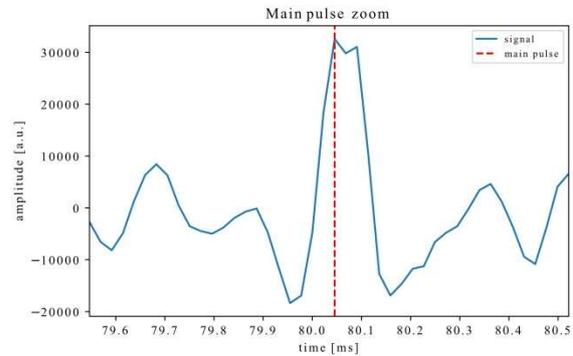

Figure 3: Zoom on main pulse (OK)

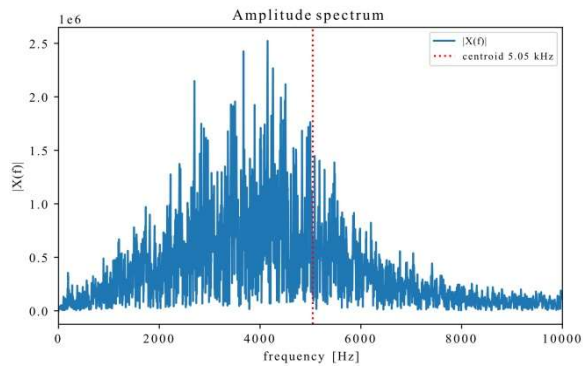

Figure 4: Amplitude spectrum (OK)

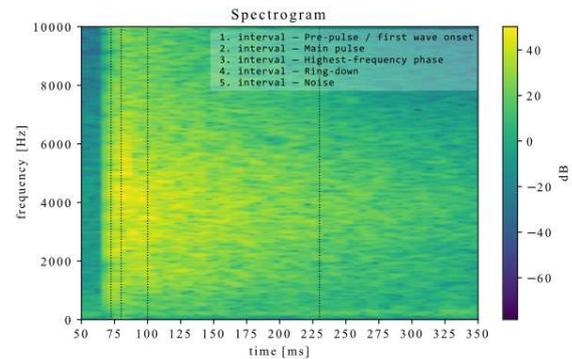

Figure 5: Spectrogram (OK)

| Interval [ms] | Physical process | Information |
|:---:|:---:|:---:|
| 0-72 | noise before the captured signal | no information content |
| 72-81 | shock wave + emission into air | main pulse energy |
| 81-101 | resonance + onset of cavitation | highest frequencies |
| 101-230 | cavitation + chamber reflections | reverberation |
| >230 | room reverberation | low information content |

Table 1: Time intervals of the acoustic signal and their physical interpretation.

# 4. Feature Extraction and Model Training

For each recorded pulse, features describing its overall energy, frequency characteristics, and time-resolved behaviour were extracted. Global characteristics (see Table 2) include the RMS value of the signal, the spectral centroid, and the ratio of high-frequency energy above 6 kHz. The time-frequency behaviour of the pulse was further analysed by detecting secondary maxima in the signal envelope and by evaluating the relative energy and spectral structure across several consecutive time intervals (see Table 3). This set of features provides a complete description of the acoustic response of individual pulses and serves as the input representation for subsequent classification of LSP quality.

| Feature | Description | Interpretation |
| --- | --- | --- |
| *rms_global* | RMS amplitude value | Total pulse energy |
| *centroid_global_hz* | Spectral centroid of the signal | Higher → stronger cavitation |
| *hf_ratio_global* | Fraction of energy above 6 kHz | Sharpness of the shock wave |
| *n_secondary_300_1000* | Number of secondary peaks | Cavitation / acoustic reflections |

Table 2: Global acoustic features.

| Feature | Description | Interpretation |
| --- | --- | --- |
| *rms_rel_int1-5* | Relative energy in intervals | Ratio of pulse energy vs. reverberation |
| *hf_ratio_int1-5* | Fraction of HF energy in intervals | Distinguishing sharp shocks from decay |

Table 3: Interval features after the main pulse.

**Data split and model training.** The extracted features were used to train a Random Forest model. Given the behaviour of the data and the limited number of samples, this model is more suitable than classification using CNNs on spectrograms, which would require a much larger volume of data. The dataset was split into training (60%), validation (20%), and test (20%) subsets. The model was trained with 300 decision trees and class-balanced weighting, which increases robustness. The model was first trained on the training set and validated on the validation data, where it achieved 100% accuracy. It was then further trained on the combination of the training and validation sets and finally tested on the previously unseen test set with repeated evaluation.

## 5. Results interpretation

Analysis of feature importance showed that the model bases its decisions primarily on global signal parameters (see Fig. 6) and on the time-frequency evolution of the signal. This finding is consistent with practical experience, where an operator can distinguish OK and NOT OK pulses based on audible differences in frequency content.

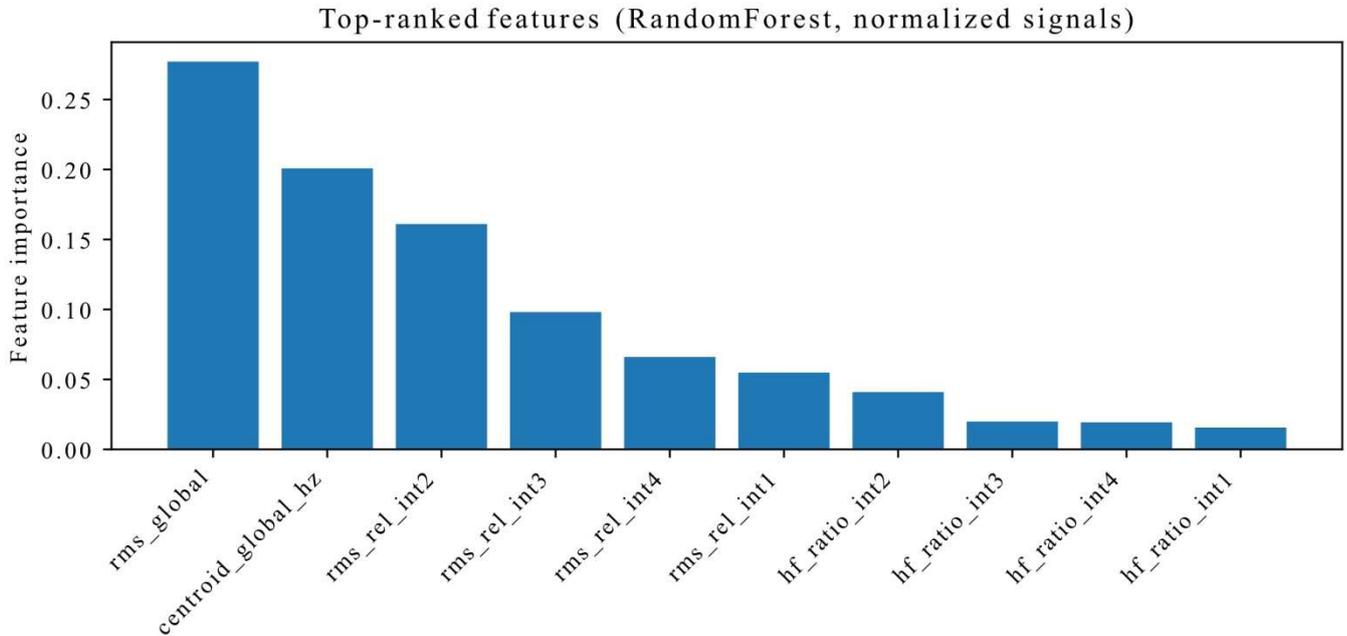

Figure 6: Discrimination power of acoustic metrics.

The most important feature was the total acoustic energy (RMS global) (see Fig. 7), which was typically higher for defective pulses. We attribute this to the absence of damping by the water layer. Conversely, a higher spectral centroid (centroid global Hz) and a higher fraction of high-frequency energy shortly after the main pulse (hf ratio int2) were characteristic of OK pulses, which generate a sharper response. We attribute these higher frequencies to cavitation [10,13,14].

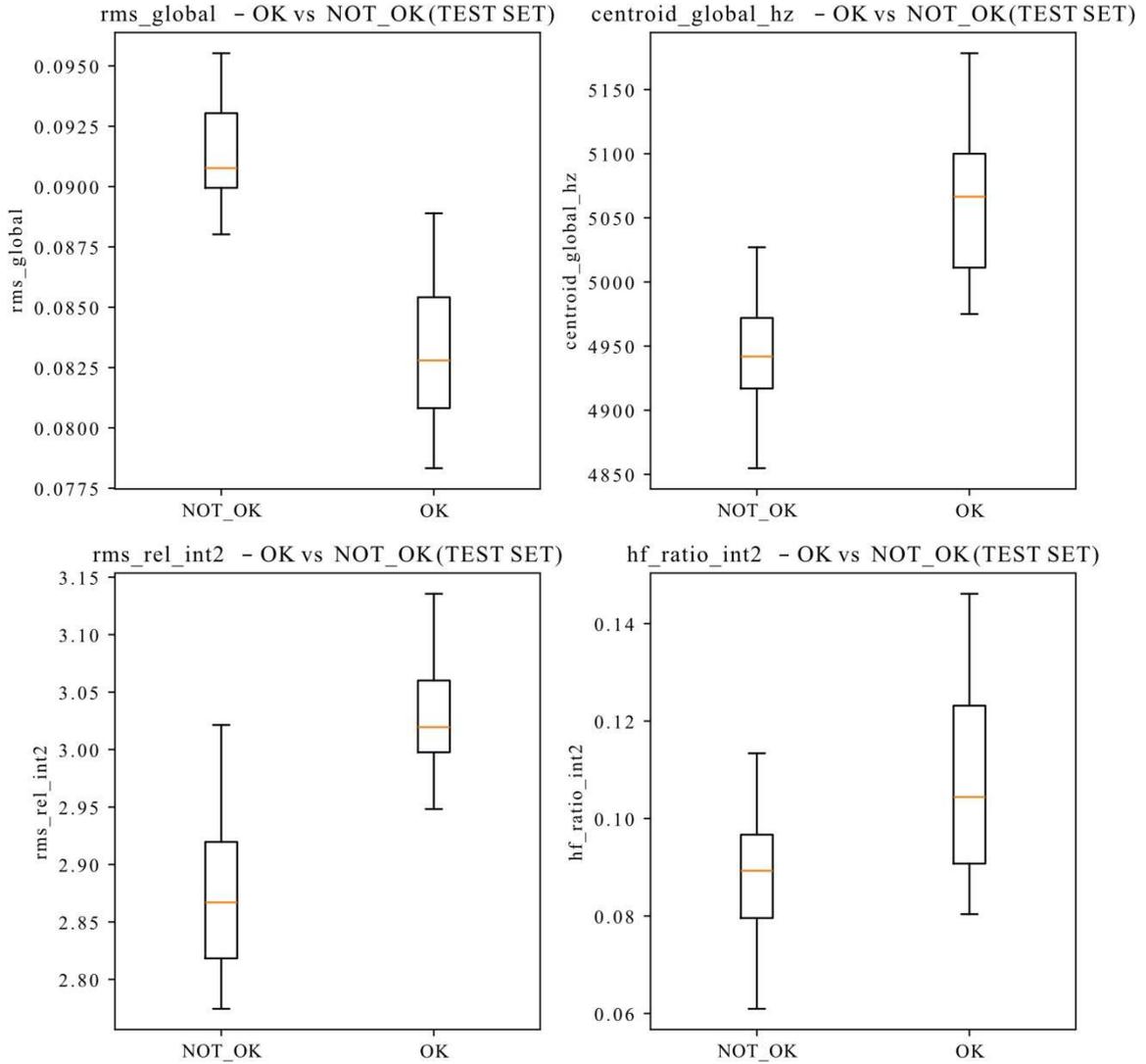

Figure 7: Boxplots of key acoustic features for *OK* vs. *NOT OK* pulses.

## 6. Discussion

This work demonstrates that the acoustic response captured by a low-cost microphone is sufficient for automatic classification of the quality of individual LSP pulses into *OK*/*NOT OK* [9–17]. The approach is attractive in practice because it enables low-cost, fast and objective in-line inspection without the need for destructive residual stress measurements. The observed differences between the classes are likely related to the role of the water layer, which affects damping and cavitation, and thus changes the total signal energy and the proportion of high-frequency components [10–14]. The results are consistent with current trends in on-line monitoring of LSP, but shift them towards a simpler and more easily integrated variant [15–19].

The transferability of the concept is promising, but will depend on the material, geometry, acoustic environment and microphone position. It should also be noted that in this study the classes are defined by a strongly different process condition (with a water layer vs. without water) and the dataset is currently limited to a single configuration, which may simplify the separation. In practice, the method can be used as a simple alarm for detecting incorrect pulses during the process. A logical next step is to validate the method on more materials and finer types of process deviations, while including greater variability of the measurement setup [8,19,20].

## 7. Conclusion

Our results show that the quality of individual LSP pulses can be effectively quantified using a standard microphone, without the need for destructive testing, **with the proposed model achieving 100% success rate**. The proposed model can be integrated directly into the control system of the laser process, where it can serve as a real-time diagnostic tool.

The method is based on physically meaningful phenomena and provides a transparent basis for further analysis and extension. This approach therefore has the potential to increase process reliability and facilitate the industrial deployment of LSP without subsequent expensive quality-control procedures.

**Acknowledgements:** This work was co-funded by the European Union and the state budget of the Czech Republic under the project LasApp CZ.02.01.01/00/22_008/0004573.

**Conflicts of Interest:** The authors declare no conflicts of interest.

**Data availability statement:** The datasets used and/or analyzed during the current study are available from the corresponding author on reasonable request.

**Author contribution statement:** Project Management: J.Kočí, Mechanical Engineering: I.Mašín, Sensorics: M.Kotek, Control Systems: T.Martinec.